\documentclass[twocolumn,english,pre]{revtex4-2}
\usepackage[T1]{fontenc}
\usepackage[latin9]{inputenc}
\usepackage{amsmath}
\usepackage{graphicx}
\usepackage{color}
\makeatletter
\usepackage{hyperref}

\makeatother

\usepackage{babel}

\newcommand{\dif}{{\rm d}}
\newcommand{\bk}{{\bf k}}
\newcommand{\br}{{\bf r}}
\newcommand{\bka}[1]{\left\langle #1 \right\rangle}
\newcommand{\bks}[1]{\left[ #1 \right]}
\newcommand{\bkr}[1]{\left( #1 \right)}

\begin{document}
\title{Unraveling on Kinesin Acceleration in Intracellular Environments:
A Theory for Active Bath }
\author{Mengkai Feng}
\author{Zhonghuai Hou}
\email{E-mail: hzhlj@ustc.edu.cn}

\affiliation{Hefei National Research Center for Physical Sciences at the Microscale
\& Department of Chemical Physics, University of Science and Technology
of China, Hefei, Anhui 230026, China}
\date{\today}
\begin{abstract}
Single molecular motor kinesin harnesses thermal and non-thermal fluctuations to transport various cargoes along microtubules, converting chemical energy to directed movements. 
To describe the non-thermal fluctuations generated by the complex environment in living cells, we establish a bottom-up model to mimic the intracellular environment, by introducing an active bath consisting of active Ornstein-Uhlenbeck (OU) particles.
Simulations of the model system show that kinesin and the probe attached to it are accelerated by such active bath. 
Further, we provide a theoretical insight into the simulation result by deriving a generalized Langevin equation (GLE) for the probe with a mean-field method, wherein an effective friction kernel and fluctuating noise terms are obtained explicitly.
Numerical solutions of the GLE show very good agreement with simulation results. 
We sample such noises, calculate their variances and non-Gaussian parameters, and reveal that the dominant contribution to probe acceleration is attributed to noise variance.
\end{abstract}
\maketitle

\section{Introduction}
Kinesins are a class of molecular motor proteins that are driven by hydrolysis of adenosine triphosphate (ATP) and move along microtubule filaments to transport various cargos \cite{berg2002biochemistry,hirokawa2009kinesin,vale2003molecular}.
The kinetic mechanism of kinesin movement has been well studied through single-molecule measurement technologies \cite{milic2014kinesin, dogan2015kinesin, isojima2016direct}.
Beyond direct ATP propulsion, in living cells, cargo-loaded kinesin utilizes thermal fluctuations to make directed motions \cite{vale1990protein, vale2003molecular}.
Besides, metabolic activities, which are hard to recur in experimental conditions (in vitro) but do occur in living cells, generate non-thermal fluctuations through energy input \cite{GUO2014cell,PARRY2014Cell,Kenji2017SciAdv,Fodor_2015,19JPCL_anomalDyn}.
A few works showed that active fluctuations have non-Gaussian properties in various physical systems, such as active swimmer suspensions \cite{kurihara2017non,Esparza2019SoftMatt,Zaid2016PRL} and cytoskeleton networks \cite{Shi2019PNAS}. Effects of these active fluctuations have become a hot topic recently in biophysics and non-equilibrium statistical physics community\cite{ariga2021noise,ariga2020experimental, kurihara2017non,ariga2018nonequilibrium}, and direct measurement of kinesin with non-thermal noises has been achieved experimentally (in vitro) \cite{ariga2021noise,ariga2020experimental,21JPCL_ECP}.

It has been shown that active fluctuations promote the transport of molecular motors as far as we know \cite{Fodor_2015,kurihara2017non, Shi2019PNAS, ezber2020dynein,ariga2020experimental,21JPCL_ECP,ariga2021noise}.
Ariga \textit{et al} \cite{ariga2021noise} studied the noise-induced acceleration of kinesin with experiments and a phenomenological theory. 
They found that kinesin accelerates under a semi-truncated L\'evy noise, and when a large hindering force is loaded, this acceleration becomes more significant. 
They also pointed out that the efficiency of kinesin is surprisingly low in vitro \cite{ariga2018nonequilibrium} so that they hypothesized the kinesin movement is likely to be optimized for noisy intracellular environment but not necessarily for extracellular situations. 
Similarly, another class of motor proteins, dynein, also exhibits analogous behavior. Ezber \textit{et al} \cite{ezber2020dynein} found that dynein harnesses active fluctuations for faster movement experimentally, and described this phenomenon with a racket potential model based on Arrhenius theory. Analogously, Pak \textit{et al} \cite{21JPCL_ECP} studied probe transport and diffusion enhancement in the ratchet potential and the presence of ``exponentially correlated Poisson (ECP) noise'' experimentally. They found that the probe velocity not only increased with noise strength, but also reached maximum for a characteristic correlation time scale and non-Gaussian distribution of such noise.

On the other hand, when biological swimmers or artificial self-propelled particles are suspended in the fluid, the transport properties of the probe can be dramatically altered. 
This constitutes a model called ``active bath'' or ``active suspension'' that has been widely investigated experimentally and theoretically in recent decades \cite{2000_PRL_Wu_bacteria_bath, 2004_PF_Kim_EnhancedDiff, 2009_PRL_Goldstein_EnhancedTrcDiff, 11sm_bacBath, lagarde2020colloidal, 2016_NatPhys_HeatEngineBacteriaBath,2017_PRE_TracerDiffActBath,2017_SciRep_Maggi_MemLessResponseAndFDT, 2020_PRL_YangMC, kanazawa2020loopy, granek2022anomalous}. 
In particular, significant progress has been made in recent years in modelling and theoretical researches, which are based on various theoretical methods, including density functional theory \citep{2007_JCP_DDFT}, non-equilibrium
linear response theory 
\citep{2017_PRE_TracerDiffActBath,2009_PRL_Maes,2013_PRE_Maes,2011_JSM_Maes_FluRes,2016_JPCM_Maes_Langevin,20_PRL_Maes,20_FrontPhys_Maes_respon},
mean-field theory method (including our previous work on the effective mobility and diffusion of a passive tracer in the active bath \citep{21FengArxiv}) 
\citep{2014_NJP_Demery_GLEforDrivenTracer,2011_PRE_Demery_TracerInFluids,2019_JSM_Dean_DrivenProbeInHarmoConfi,2019_PRL_Demery_SSPinHarmoTrap,2010_PRL_Demery_DragForce,maitra2020enhanced}, and even mode-coupling theory \citep{2013_PRE_Fuchs,21SM_VoigtmannMCT}.
The ``active bath'' model brings an available tool to investigate the probe properties in complex fluids which are far from equilibrium and evolve complicated interactions, such as cytoplasm in living cells.  
All these works inspire us to build a bottom-up model for kinesin in an intracellular environment and derive a corresponding theory that serves as a novel fundamental way to decode the kinesin acceleration in non-equilibrium
situations.

In the present work, we introduce an active bath model to mimic the cytoplasmic environment, by utilizing soft colloidal particles (also known as ``active crowder'') to imitate various proteins or vesicae, and particle activity to simulate metabolic processes. Then we investigate the effects of thermal/non-thermal fluctuation generated by these crowders on kinesin transport. Our model briefly captures the most significant parts of the system and allows a wide range of
parameters to include various kinds of situations. It brings a novel, quantifiable research approach to active fluctuations in living cell.

\begin{figure}
\begin{center}
\includegraphics[width=8.6cm]{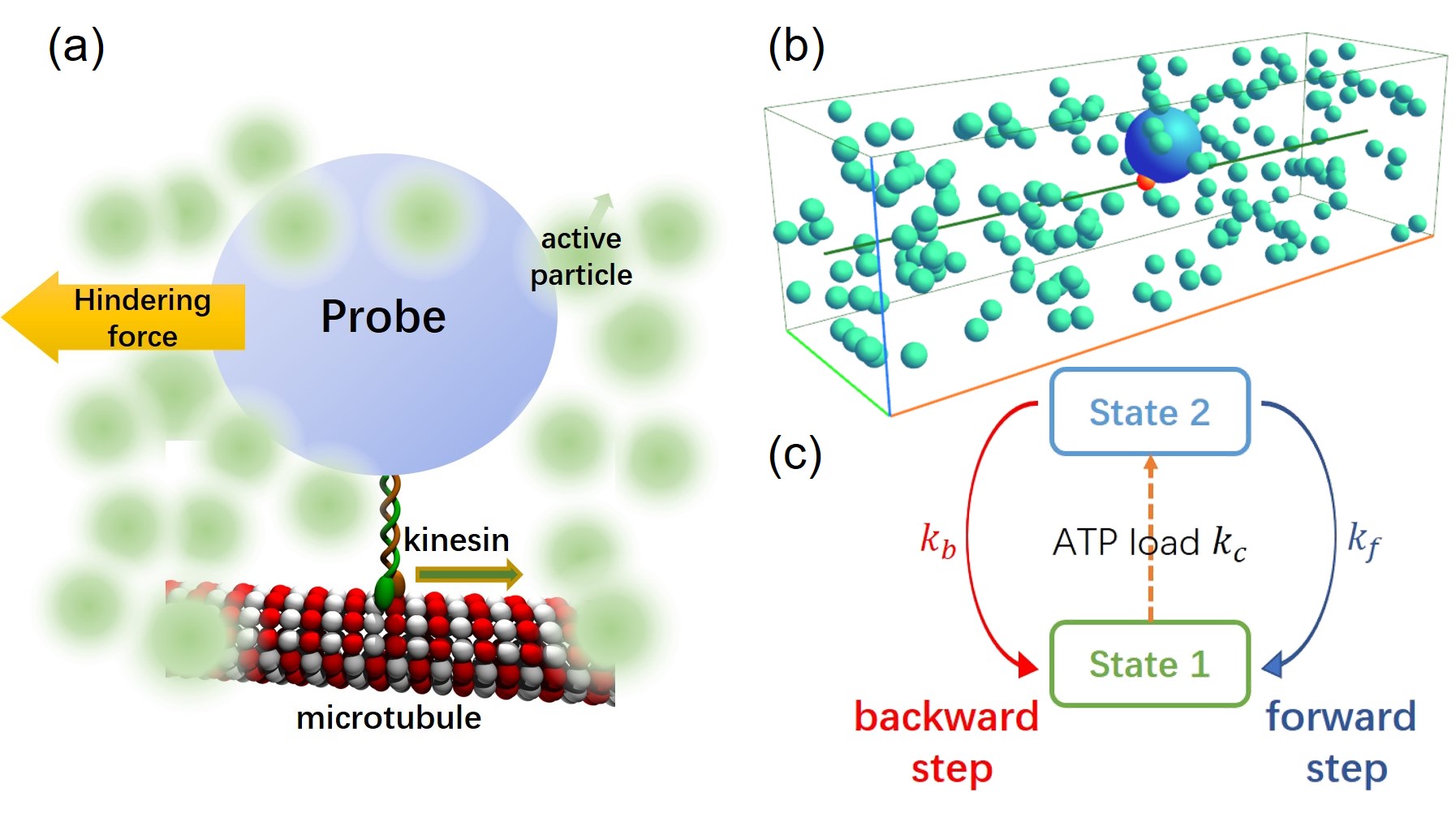} 
\includegraphics[width=8.6cm]{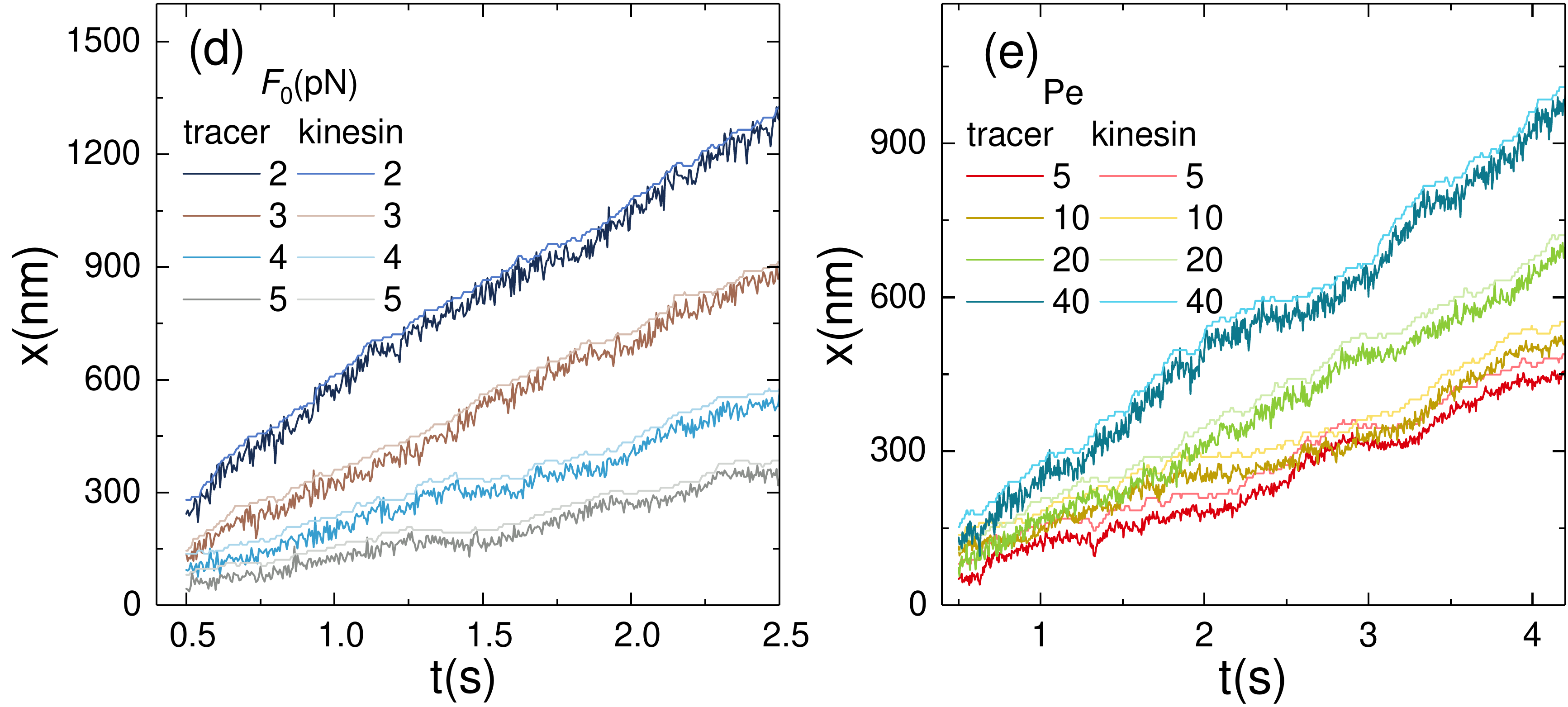}
\end{center}
\caption{(a)Cartoon for the model system; (b)Schema for actual simulation system:
large blue ball stands for the probe, small cyan balls indicate active
particles(only a few particles are shown), and red dot indicates the
kinesin;(c)Schematic diagram for Markovian model of kinesin movement.
Typical trajectories of tracer and kinesin in active bath, (d)for
different hindering forces from $-2$ to $-4$ pN with constant activity
${\rm Pe}=40$, (e) for different activity of bath under constant
drag force $F_{0}=-4$pN. }
\label{fig:schema}
\end{figure}

\section{Modeling and Simulations}
Let us consider a three-dimensional
system shown in Fig.\ref{fig:schema}(a), where a probe (or called
tracer elsewhere) attached to a kinesin is suspended in an active
bath consisting of $N$ self-propelled particles inside a box of side
length $L_{x},L_{y},L_{z}$ with periodic boundaries. These bath particles
are propelled by independent OU noises, forced by inter-particle repulsive
potentials and background thermal noises. The movement of bath particles
is governed by overdamped Langevin equations \begin{subequations}
\begin{align}
\dot{{\bf r}}_{i}= & -\mu_{b}\nabla_{i}\left[\sum_{j\neq i}V(|{\bf r}_{i}-{\bf r}_{j}|)+U(|{\bf r}_{i}-{\bf x}_{p}|)\right]\nonumber \\
 & +{\bf f}_{i}+\sqrt{2\mu_{b}k_{B}T}\boldsymbol{\xi}_{i}\\
\tau_{b}\dot{{\bf f}}_{i}= & -{\bf f}_{i}+\sqrt{2D_{b}}\boldsymbol{\zeta}_{i}
\end{align}
\label{eq:LE_bath}
\end{subequations} \\
\noindent where ${\bf r}_{i}$ is the position for $i$-th bath particle,
$\mu_{b}$ is the mobility, ${\bf x}_{p}$ is the position of the
probe particle, $V(r)$ and $U(r)$ are interacting potentials between
bath-bath particles and bath-probe respectively, ${\bf f}_{i}$ is
the propulsion force acting on $i$-bath particle with persistent
time $\tau_{b}$ and strength $D_{b}$, $k_{B}$ is Boltzmann constant
and $T$ is the background temperature, $\boldsymbol{\xi}_{i}$ and
$\boldsymbol{\eta}_{i}$ are independent Gaussian white noise vectors
in 3d space, with zero means and delta correlations $\left\langle \boldsymbol{\xi}_{i}(t)\boldsymbol{\xi}_{j}(t')\right\rangle =2\delta_{ij}\delta(t-t'){\bf I}$
and $\left\langle \boldsymbol{\zeta}_{i}(t)\boldsymbol{\zeta}_{j}(t')\right\rangle =2\delta_{ij}\delta(t-t'){\bf I}$,
where ${\bf I}$ is the unit matrix.

The molecular motor is described by a phenomenological Markov-like
kinetic diagram based on experimental observations \cite{ariga2018nonequilibrium},
wherein the complex kinesin walking process is simplified to a two-state
Markov transition. In this model, the central ATP hydrolysis and walking
process is divided into three transition steps(see Fig.\ref{fig:schema}(c)).
The first step is ATP load with constant rate $k_{c}$ and causes
a ``state transition'' (state 1 to state 2). This rate is dependent
on the concentration of ATP, and independent of any mechanical issues.
The second and third steps are mechanical transitions for forward
and backward steps with constant step size $d=8$nm along the microtubule
as well as rates $k_{f}$ and $k_{b}$ respectively. Meanwhile the
state transition accompanies both steps, from state 2 to state 1.
These two rates have both force $F$ dependent as Arrhenius-type 
\begin{equation}
k_{\{f,b\}}(F)=k_{\{f,b\}}^{0}\exp\left(\frac{d_{\{f,b\}}F}{k_{B}T}\right)
\label{eq:kfbF}
\end{equation}
where $k_{\{f,b\}}^{0}$ is the rate constant without any external
force load, $d_{\{f,b\}}$ is the characteristic distant, and all
of these parameters are fitted by experimental data. Mathematically,
the evolution of the probability of each state ($P_{1}$ and $P_{2}$)
obeys a Master equation 
\begin{equation}
\frac{{\rm d}}{{\rm d}t}P_{2}=k_{c}P_{1}-(k_{f}+k_{b})P_{2}
\label{eq:dtP2}
\end{equation}
This equation establishes the relationship between mean velocity and
all fitting parameters for kinesin systems, $\bar{v}=d\frac{(k_{f}-k_{b})k_{c}}{k_{f}+k_{b}+k_{c}}$,
which is used to identify fitting parameters mentioned above and can
be determined by experiments \cite{ariga2021noise}.

One of the most concerned quantities in our model is the position of the probe ${\bf x}_{p}$. The probe is dragged by a constant hindering force ${\bf F}_{0}$ (to mimic optical tweezers in experiments) and pulled by a molecular motor kinesin via a linear spring with stiffness $K$. To illustrate the setup, we draw a cartoon in Fig.\ref{fig:schema}(a), and show the actual simulation system in (b) wherein the kinesin and probe are both constrained to move along $\vec{e}_{x}$ direction.
The movement of the probe is also described by an overdamped Langevin equation
\begin{equation}
\dot{x}_{p}=\mu_{p}\left[K(x_{m}-x_{p})+F_{0}+F_{bath}\right]+\sqrt{2\mu_{p}k_{B}T}\xi_{t}\label{eq:LE_probe}
\end{equation}
where $x_{p}$ and $x_{m}$ are the position of the probe and the motor along $\vec{e}_x$ direction respectively, $\mu_{p}$ is the mobility of the probe, and $F_{bath}=-\frac{\partial}{\partial x_{p}}\sum_{i}U(|{\bf r}_{i}-{\bf x}_{p}|)$
is the interactions between the probe and bath particles.

For easier comparison with the previous experimental results, in simulations
we use SI unit and set $k_{B}T=4.115{\rm pN\cdot nm}$ for room temperature. Considering the intracellular environment is dense, and interactions
of various components such as proteins and vesicae are soft, we roughly set the active crowder diameter $R_{b}=160$nm and mobility $\mu_{b}=1.0\times10^{5}{\rm nm/(pN\cdot s)}$, set the bath particle density $\rho=N/(L_{x}L_{y}L_{z})=1.0/R_{b}^{3}$ as a relatively high value, choose harmonic potential as the interactions between particles, $U(r)=\frac{\kappa}{2}(\sigma_{pb}-|r|)^{2}$ for $|r|<\sigma_{pb}$ and $V(r)=\frac{\kappa}{2}(\sigma_{bb}-|r|)^{2}$ for $|r|<\sigma_{bb}$ , where $\sigma_{pb}=(R_{p}+R_{b})/2=340$nm, $\sigma_{bb}=R_{b}=160$nm is the interacting distance of probe-bath particles and bath-bath particles, $\kappa$ is the interaction strength which is set as a constant. Other parameters and simulation details are shown in App.\ref{sec:ns}. 
In this work, the main control parameters are the activity of active crowder, measured by P\'eclet number, which is dimensionless and defined as ${\rm Pe}=\frac{\sqrt{D_{b}/\tau_{b}}R_{b}}{\mu_{b}k_{B}T}$, where $\sqrt{D_{b}/\tau_{b}}$ is standard deviation of ${\bf f}_{i}$,
as well as the persistent time of active crowder $\tau_{b}$. 

Figure\ref{fig:schema}(d,e) shows several typical simulation trajectories
of the kinesin and the probe attached to it. Due to the kinesin walking
process, all kinesin/probe moves toward positive $x$ direction. With
the constant bath activity and kinetics parameters of kinesin, the
influence of hindering load force on kinesin/probe movement is shown
in Fig.\ref{fig:schema}(d). As a matter of course, larger load force
leads to slower movement, as well as larger distance between kinesin
and probe. Besides, the active fluctuations on probe contribute significant
promotion effect. As shown in Fig.\ref{fig:schema}(e), with the constant
hindering force, larger bath activity induces faster kinesin/probe
movement. 

\begin{figure}
\centering{}\includegraphics[width=8cm]{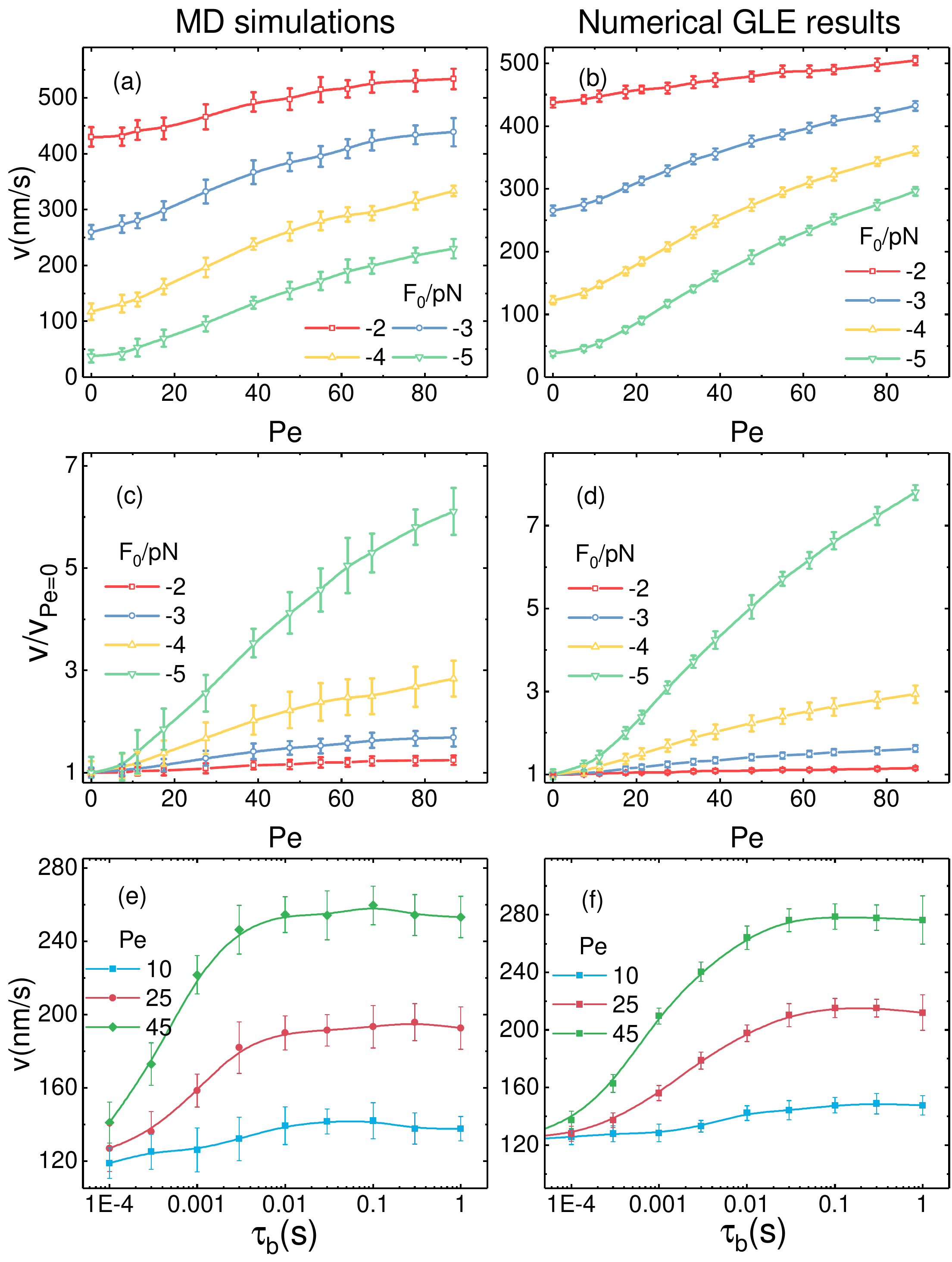}\caption{(a)Simulation results, average velocity of the probe in the active
bath of activity. Each marker indicates various average loads from
$-2$ to $-5$pN; (b)Numerical solution of GLE with the same conditions
and parameters as (a); (c,d)Relative velocity of the probe under the
same parameters with (a,b) respectively; (e,f)Probe velocity for various
activity Pe, figures are the plots of $v$ as the function of persistent
time $\tau_{b}$. Other parameters herein: for (a)-(d), we use $\tau_{b}=0.01{\rm s}\approx\frac{R_{b}^{2}}{6\mu_{b}k_{B}T}$,
which is the characteristic rotational time of a Brownian particle
with diameter $R_{b}$. All figures use $\kappa=0.003{\rm pN/nm}$
that indicates a weak interaction, we also test other values of $\kappa$,
qualitatively results are not affected by this parameter.}
\label{fig:v_Pe}
\end{figure}

Average velocities of probe $v$ for variant bath activities are shown
in Fig.\ref{fig:v_Pe}(a) and (c) for a normalized version, and each
marker indicates the hindering force $F_{0}$ from $-2$ to $-5$pN.
Results show that probe velocity $v$ increases with bath activity
Pe monotonically in all cases. Especially, normalized velocity $v/v_{{\rm Pe=0}}$
shows a stronger enhancement under high hindrance loads. This result
is very similar to a most recent \textit{in vitro} experiment \cite{ariga2021noise},
wherein the researchers have used optical tweezers to apply a ``semitruncated
L\'evy noise'' and an additional constant load force on the probe.
They found that motor/probe velocity increases with the magnitude
of the noise, and that such increases are larger for the stronger
load forces. We also investigate the kinesin velocity dependence on
persistent time of active bath particles with fixed activity Pe, shown
in Fig.\ref{fig:v_Pe}(e). Simulations show that the probe velocity
increases with persistent time $\tau_{b}$ at first and next reaches
a platform. Then, probe velocity weakly decreases at large $\tau_{b}$
region. 

\section{Theory of Active Bath}
To understand our simulation results, we develop a mean-field theory method to investigate the system theoretically.
The starting point of the theory is the overdamped Langevin equations
\eqref{eq:LE_bath}, and the objective of the theory is to obtain
an effective movement equation that only contains probe and kinesin
variables. To eliminate numerous degrees of freedom of bath particles,
we describe the model system at a coarse-grained level, employing
an evolution equation for bath particles' density profile $\rho({\bf r},t)$
\begin{align}
\frac{\partial\rho({\bf r},t)}{\partial t}= & \mu_{b}\nabla_{{\bf r}}\cdot\rho({\bf r},t)\nabla_{{\bf r}}\Big[\int\rho({\bf r}',t)V(|{\bf r}-{\bf r}'|){\rm d}{\bf r}'\nonumber \\
 & +U(|{\bf r}-{\bf x_{p}}|)\Big]+\nabla\cdot\left[\sqrt{\rho({\bf r},t)}\boldsymbol{\xi}^{A}({\bf r},t)\right]\\
 & +\nabla\cdot\left[\sqrt{\rho({\bf r},t)}\boldsymbol{\xi}^{T}({\bf r},t)\right]+\mu_{b}k_{B}T\nabla^{2}\rho({\bf r},t)\nonumber 
\end{align}
which is a Dean-like equation for active particle system, wherein
$\boldsymbol{\xi}^{A,T}\left({\bf r},t\right)$ are noise filed functions.
To embody the effect of such density profile on probe movement, we
firstly solve this equation in Fourier space formally, 
\begin{align}
\frac{\partial\rho_{k}(t)}{\partial t} & \approx-\mu_{b}k^{2}\left[(k_{B}T+\rho V_{k})\rho_{k}(t)+\rho U_{k}e^{i{\bf k}\cdot{\bf x}_{p}}\right]\nonumber \\
 & +i\sqrt{\rho}{\bf k}\cdot\left[\tilde{\boldsymbol{\xi}}^{T}({\bf k},t)+\tilde{\boldsymbol{\xi}}^{A}({\bf k},t)\right]
\end{align}
where $\rho$ is the number density of bath particle, $\tilde{\boldsymbol{\xi}}^{A,T}({\bf k},t)$,
$U_{k},V_{k}$ are Fourier transform of noises $\boldsymbol{\xi}^{A,T}({\bf r},t)$
and potentials $U(r),V(r)$ respectively, with time correlations $\left\langle \tilde{\xi}_{\alpha}^{A*}({\bf k},t)\tilde{\xi}_{\beta}^{A}({\bf k}',t')\right\rangle =\frac{D_{b}}{\tau_{b}}\delta_{\alpha\beta}(2\pi)^{3}\delta({\bf k}-{\bf k}')e^{-|t-t'|/\tau_{b}}$
and $\left\langle \tilde{\xi}_{\alpha}^{T*}({\bf k},t)\tilde{\xi}_{\beta}^{T}({\bf k}',t')\right\rangle =2\mu_{b}k_{B}T\delta_{\alpha\beta}(2\pi)^{3}\delta({\bf k}-{\bf k}')\delta(t-t')$.
Then insert this formal solution into Eq.(\ref{eq:LE_probe}) by utilizing
an identity $-\nabla_{{\bf x}_{p}}\sum_{i}U(|{\bf r}_{i}-{\bf x}_{p}|)\equiv\frac{1}{(2\pi)^{3}}\int i{\bf k}e^{-i{\bf k}\cdot{\bf x}_{p}}\rho_{k}(t) U_k {\rm d}^{3}{\bf k}$.
After some appropriate approximations, we obtain a generalized Langevin
equation for the probe
\begin{align}
\dot{x}_{p}(t)= & -\mu_{p}\int_{-\infty}^{t}\zeta(t-s)\dot{x}_{p}(s){\rm d}s+\eta_{A}(t)+\eta_{T}(t)\nonumber \\
 & +\mu_{p}[K(x_{m}-x_{p})+F_{0}]+\sqrt{2\mu_{p}k_{B}T}\xi_{t}\label{eq:GLE_probe}
\end{align}
with memory kernel 
\begin{equation}
\zeta(t)=\frac{\mu_{p}\mu_{b}\rho}{3(2\pi)^{3}}\int k^{4}U_{k}^{2}a_{k}e^{-t/a_{k}}{\rm d}^{3}{\bf k}
\end{equation}
where $a_{k}=\left[\mu_{b}k^{2}(k_{B}T+\rho V_{k})\right]^{-1}$ is
a characteristic time scale, and $\eta_{A,T}$ are complicated colored
noise 
\begin{align}
\eta_{A,T}(t)= & \frac{\mu_{p}\sqrt{\rho}}{(2\pi)^{3}}\int ik_{x}U_{k}\nonumber \\
 & \times\int_{-\infty}^{t}e^{-(t-s)/a_{k}}{\bf k}\cdot\tilde{\boldsymbol{\xi}}^{A,T}({\bf k},s){\rm d}s{\rm d}^{3}{\bf k}\label{eq:etaAT}
\end{align}
with time correlation functions\begin{subequations}
\begin{align}
\left\langle \eta_{T}(t)\eta_{T}(t')\right\rangle  & =\frac{\mu_{t}^{2}\rho\mu_{b}k_{B}T}{3(2\pi)^{3}}\int U_{k}^{2}k^{4}a_{k}e^{-|t-t'|/a_{k}}{\rm d}^{3}{\bf k}\\
\left\langle \eta_{A}(t)\eta_{A}(t')\right\rangle  & =\frac{\mu_{t}^{2}\rho D_{b}}{3(2\pi)^{3}}\int U_{k}^{2}k^{4}\frac{1}{(\tau_{b}/a_{k})^{2}-1}\nonumber \\
 & \times\left[\tau_{b}e^{-|t-t'|/\tau_{b}}-a_{k}e^{-|t-t'|/a_{k}}\right]{\rm d}^{3}{\bf k}
\end{align}
\label{eq:corr_eta}
\end{subequations}Herein, a generalized fluctuation-dissipation relationship (FDR) is
reveal between memory kernel $\zeta(t)$ and noise $\eta_{T}$, and
the OU noise of the bath particle brings an explicitly violation of
the FDR. When the activity of the bath is absent, Eq.(\ref{eq:GLE_probe})
reduces to a GLE in equilibrium and the FDR holds naturally.

Equations (\ref{eq:GLE_probe})-(\ref{eq:etaAT}) are main theoretical
results of the present work. They unravel the properties of noise
generated by active environment, and allow us to directly calculate
the probe movement and average velocity. Numerical solutions of Eq.(\ref{eq:GLE_probe})
are shown in Fig.\ref{fig:v_Pe}(b), (d) and (f), wherein the parameters
are chosen same as (a),(c) and (e) respectively. Compared with simulation
results, the GLE reproduces the acceleration effect of active crowders
(a)-(d), quantitatively in most cases. Surprisingly, GLE solutions
also show very similar behavior of relationship between probe velocity
and persistent time $\tau_{b}$, which further confirms the non-trivial
phenomenon. 

Theoretical explanations about the mechanism of kinesin acceleration
are still in development. In Ref.\cite{ariga2021noise}, the authors
pointed out that the amplitude of noise is a major factor. Yet in
a ratchet potential model \cite{21JPCL_ECP}, not only the noise strength
significantly influence the probe dynamics, but also non-Gaussian
property and time correlation behavior of the noise. Herein, with
the help of the GLE, it is feasible to investigate which property
of the noise dominates kinesin acceleration. 

\begin{figure}
\centering \includegraphics[width=8.6cm]{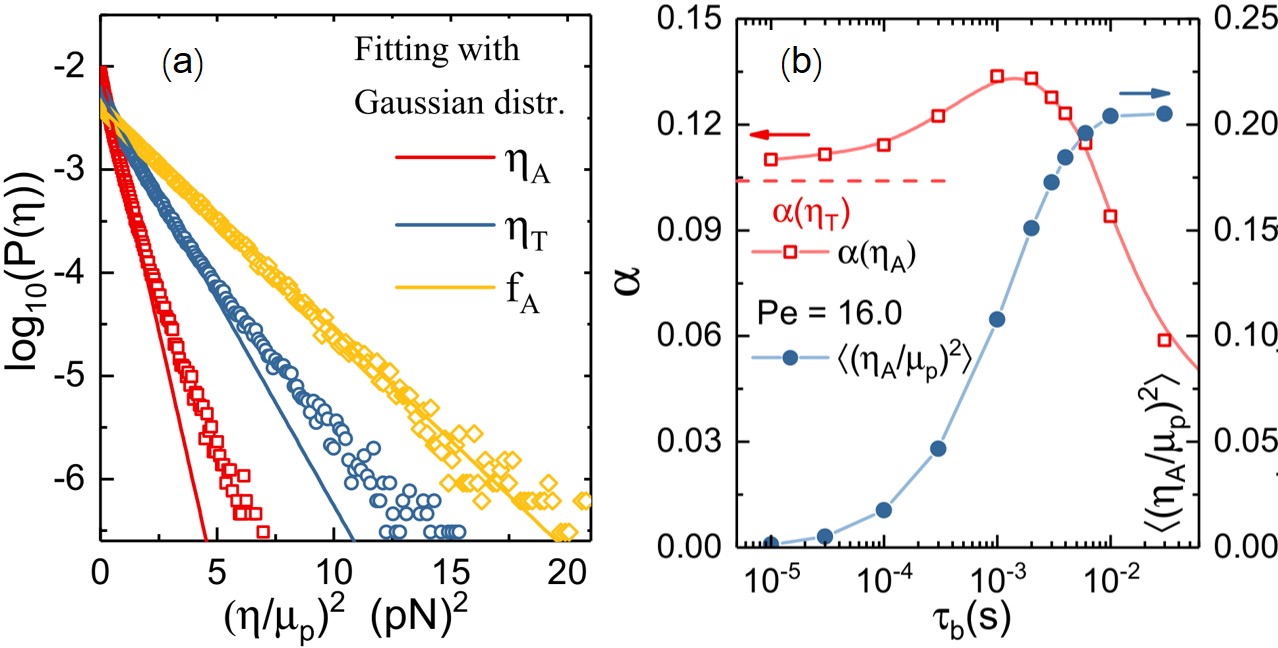}

\caption{(a) Distributions for noise $\boldsymbol{\eta}_{A}$ (red squares)
and $\boldsymbol{\eta}_{T}$ (blue round dots), solid lines are fitting
lines with hypothetical Gaussian distributions. ${\rm f}_{A}$ is
an OU process as a reference(yellow diamonds and line), which satisfies
an exact Gaussian distribution. Herein, we set $\tau_{b}=0.01{\rm s}$.
(b) Non-Gaussian parameters $\alpha$ for $\boldsymbol{\eta}_{A}$
and $\boldsymbol{\eta}_{T}$ (red squares and left vertical axis),
the formal one depends on persistent $\tau_{b}$ which is drawn here
as horizontal axis. Also, we plot the variance of noise $\boldsymbol{\eta}_{A}/\mu_{p}$
(blue points and right vertical axis). Other parameters for all subplots:
$\kappa=0.01{\rm pN/nm}$, ${\rm Pe}=16.0$.}
\label{fig:noise}
\end{figure}

Firstly, we focus on the strength (or amplitude) of the colored noise
$\eta_{A,T}(t)$. According to Eq.(\ref{eq:etaAT}), or more straightforwardly,
the time correlation function of $\eta_{A,T}(t)$, the explicit expression
for variance \begin{subequations}
\begin{align}
\left\langle \eta_{T}^{2}\right\rangle = & \frac{\mu_{t}^{2}\rho\mu_{b}k_{B}T}{3(2\pi)^{3}}\int U_{k}^{2}k^{4}a_{k}{\rm d}^{3}{\bf k},\label{eq:EeT2}\\
\left\langle \eta_{A}^{2}\right\rangle = & \frac{\mu_{t}^{2}\rho D_{b}}{3(2\pi)^{3}}\int\frac{U_{k}^{2}k^{4}a_{k}^{2}}{\tau_{b}+a_{k}}{\rm d}^{3}{\bf k},\label{eq:EeA2}
\end{align}
\end{subequations}can be obtained, therefore $\left\langle \eta_{A}^{2}\right\rangle \propto{\rm Pe}^{2}\tau_{b}\int\frac{k^{4}U_{k}^{2}a_{k}^{2}}{\tau_{b}+a_{k}}{\rm d}^{3}{\bf k}$.
As shown in Fig.\ref{fig:v_Pe}(b) and (d), probe velocity $v$ increases
with Pe monotonically when $\tau_{b}$ is constant. Although the analytical
relation between $v$ and Pe is not given due to the complexity of
memory kernel and colored noise, qualitatively variance of noise $\eta_{A}$
definitely makes a positive contribution to kinesin acceleration.

Another quantity we concerned is the non-Gaussian property of these
two colored noises. To intuitively show the distributions of such
noises, we plot the probability distribution function $P(\eta)$ in
Fig.\ref{fig:noise}(a). Red square and blue round hollow dots represent
$\eta_{A}$ and $\eta_{T}$ respectively, and solid curves are their
Gaussian fitting. Interestingly, both $\eta_{T}$ and $\eta_{A}$
show heavy tail distributions and clearly deviate from Gaussian distributions.
As a contrast, the distribution function of OU noise ${\rm f}_{A}$
is also plotted with yellow diamond dots, which perfectly satisfies
Gaussian distribution. Noticing that L\'evy noise also have such heavy
tail distribution\cite{ariga2021noise}, as well as the ECP noise\cite{21JPCL_ECP},
they all have non-trivial acceleration effect on kinesin. To quantitatively
investigate this property, we then calculate the non-Gaussian parameter
$\alpha(\eta)=\frac{\left\langle \eta^{4}\right\rangle }{3\left\langle \eta^{2}\right\rangle ^{2}}-1$
of $\eta_{A,T}(t)$. These quantities are not functions of temperature
$T$ nor activity Pe, therefore the contribution of non-Gaussian property
cannot be seen in Fig.\ref{fig:v_Pe}(a)-(d). Yet $\alpha(\eta_{A})$
is a function of $\tau_{b}$, and both simulation and GLE solution
show the same dependency relationship of kinesin velocity on $\tau_{b}$.
Hence we plot non-Gaussian parameter $\alpha$ (red squares, left axis) and corresponding noise variance (blue dots, right axis) as functions of persistent time $\tau_{b}$ in Fig.\ref{fig:noise}(b). 
When $\tau_{p}\rightarrow0$, $\eta_{A}$ reduces to the noise $\eta_{T}$ (under an effective temperature $T_{{\rm eff}}=D_{b}/(\mu_{b}k_{B})$), and its non-Gaussian parameter is shown as a red horizontal dash line in Fig.\ref{fig:noise}(b). 
As $\tau_{b}$ increases, $\left\langle \eta_{A}^{2}\right\rangle $ monotonically increases and then reaches to a plateau, which is very similar to the velocity increase with $\tau_{b}$ at short and mediate region.
As shown in Fig.\ref{fig:v_Pe}(e,f), when $\tau_{b}$ is large enough, the kinesin velocity slightly decreases with $\tau_{b}$. This weak decrease behavior  has not be seen in the noise variance. On the contrary, a strongly non-monotonic dependence of $\alpha(\eta_{A})$ on $\tau_{b}$ is observed. The non-Gaussian parameter $\alpha (\eta_A)$ rapidly decreases with $\tau_b$ when it is large. This phenomenon is very likely to lead to the weak decrease of the kinesin velocity. 
In general, variance indeed make the major contribution to the kinesin acceleration, while non-Gaussian property also makes a minor yet positive contribution to it. 

\section{Conclusion}
In summary, we build a bottom-up model consisting
of a Markovian kinesin model and an active particle bath to investigate
the acceleration behavior of kinesin and probe attached to it in complex
intracellular environment. Simulations show kinesin velocity increases
with bath activity monotonically, especially for larger load situations
where more significant acceleration effect is observed. We also establish
a coarse-grained theoretical framework to describe the active bath
and obtain a generalized Langevin equation for probe movement. The
effects of active bath on the probe are simplified into a memory kernel
and two effective noises. Numerical calculations of the GLE show very
good agreement with simulation data. Furthermore, the introduction
of the theory allows us to study the noise property conveniently and
to investigate which one of them is the essential to kinesin acceleration.
Comparing simulations and numerical solutions for GLE, we find out
that the variance of noise plays a major role in kinesin acceleration,
while non-Gaussian property brings positive yet minor contributions.

Our model and theory bring a novel, quantifiable research approach
to active fluctuations in living cells, which bridges between phenomenological
description of kinesin movement and underlying principles of statistical
physics. For further study, with more information input such as accurate
interacting parameters, we believe our model could give more accurate
results, and deeper understanding on the noise property. In addition,
the theory of active bath is independent of the kinesin model, which
also serves as a new way to investigate active environment. The generality
of which could lead to numerous other applications in other probe-bath
interacting systems.

\section{Acknowledgement}
This work is supported by MOST(2018YFA0208702) and NSFC (32090044, 21833007).

~\\
\rule[-10pt]{8cm}{0.05em}

\appendix

\section{Numerical Simulations \label{sec:ns}}
Numerical simulations are run in a three-dimensional box $ (L_x, L_y, L_z)=( 40\sigma,  10\sigma,  10\sigma)$ with periodic boundary, where $\sigma=160$nm as the unit of length. In the present coarse-grained model, both the kinesin and the probe's movements are constrained on a fixed line $(y,z)=(L_y/2, L_z/2)$. The volume repulsive interactions are only considered between bath-bath particles and bath-probe, meaning that the kinesin's volume repulsive interaction is not considered. The diameter of bath particle and the probe are set as $R_b=\sigma, R_p=3.25\sigma$, so that inter-particle distance $\sigma_{pb}=2.125\sigma$. The temperature is set as the room temperature, therefore $k_BT = 4.115 {\rm pN \cdot nm}$, which is used as the unit of the energy. The mobility of bath particle is $\mu_b=1.0\times 10^5 {\rm nm/(pN\times s)}$, which can be used to label the unit of time $\tau_u = \sigma^2 / (\mu_b k_B T)=6.22\times 10^{-2} {\rm s}$. 
We set the probe diameter $R_{p}=520$nm and mobility $\mu_{p}=0.308\times10^{5}{\rm nm/(pN\cdot s)}$.

In simulations, we use the time step $\delta t=5.0\times 10^{-6} {\rm s}$ (to keep $k_{c,b,f} \delta t \ll 1$, $\delta_t \ll \tau_b$ and $\delta t \ll \tau_u$). For each time interval, both the Markovian dynamics for kinesin and the Langevin dynamics for probe and bath particles are performed.
For each simulation, the system is allowed to reach a steady state over $10^6 \delta t$, and then the kinesin/probe's displacements and velocities are averaged over following $10^7 \delta t$ time interval. The variance of the velocity is calculated by at least 20 times simulations with the exact same parameters and different random number seeds. We find that more average counts did not have a significant effect on the reduction of the variance.

For the numerical calculation of the GLE, the time step is also set as $\delta t=5.0\times 10^{-6} {\rm s}$. The generation of the complex color noises is shown in App.\ref{sec:eta}. Velocities and their variances are calculated by over $10^7$ time steps and 50 trajectories.

The Markov transition migrates from Ref.\cite{ariga2021noise}, and parameters in Eqs.\ref{eq:kfbF} and \ref{eq:dtP2} also come from this reference: $k_f^0 = 1002{\rm s^{-1}}$, $k_b^0 = 27.9{\rm s^{-1}}$, $k_c = 102{\rm s^{-1}}$, $d_f = 3.61{\rm nm}$, $d_b = 1.14{\rm nm}$.

\begin{widetext}
\section{Dean's equation for active bath and effective generalized Langevin
equation for probe\label{sec:dean} }

This section gives the derivation details of Eq.(4) in main text.
The starting point is the Langevin equation for bath particles \begin{subequations}
\begin{align}
\dot{{\bf r}}_{i}= & -\mu_{b}\nabla_{i}\left[\sum_{j\neq i}V(|{\bf r}_{i}-{\bf r}_{j}|)+U(|{\bf r}_{i}-{\bf x}_{p}|)\right]+{\bf f}_{i}+\sqrt{2\mu_{b}k_{B}T}\boldsymbol{\xi}_{i}\label{eq:LE_bath-1}\\
\tau_{b}\dot{{\bf f}}_{i}= & -{\bf f}_{i}+\sqrt{2D_{b}}\boldsymbol{\eta}_{i}
\end{align}
\end{subequations}Introducing the single particle density $\rho_{i}({\bf r},t)=\delta({\bf r}-{\bf r}_{i}(t))$
and the collective one $\rho({\bf r},t)=\sum_{i=1}^{N}\rho_{i}({\bf r},t)$,
for an arbitrary function of bath particle coordinate $g({\bf r}_{i})$
with natural boundary condition, according to the It\=o calculus, one
has 
\begin{align}
\frac{{\rm d}g({\bf r}_{i})}{{\rm d}t}= & \left\{ -\mu_{b}\nabla_{i}\left[\sum_{j\neq i}V(|{\bf r}_{i}-{\bf r}_{j}|)+U(|{\bf r}_{i}-{\bf x_{p}}|)\right]+{\bf f}_{i}+\sqrt{2\mu_{b}k_{B}T}\boldsymbol{\xi}_{i}\right\} \cdot\nabla_{i}g({\bf r}_{i})\nonumber \\
 & +\mu_{b}k_{B}T\nabla_{i}^{2}g({\bf r}_{i})\nonumber \\
= & \int\rho_{i}({\bf r},t)\Bigg\{\left(-\mu_{b}\nabla_{{\bf r}}\left[\sum_{j\neq i}V(|{\bf r}-{\bf r}_{j}|)+U(|{\bf r}-{\bf x_{p}}|)\right]+{\bf f}_{i}+\sqrt{2\mu_{b}k_{B}T}\boldsymbol{\xi}_{i}\right)\cdot\nabla_{{\bf r}}g({\bf r})\nonumber \\
 & +\mu_{b}k_{B}T\nabla_{{\bf r}}^{2}g({\bf r})\Bigg\}{\rm d}{\bf r}\nonumber \\
= & \int\rho_{i}({\bf r},t)\Bigg\{\left[-\mu_{b}\nabla_{{\bf r}}\left[\int\rho({\bf r}',t)V(|{\bf r}-{\bf r}'|){\rm d}{\bf r}'+U(|{\bf r}-{\bf x_{p}}|)\right]+{\bf f}_{i}+\sqrt{2\mu_{b}k_{B}T}\boldsymbol{\xi}_{i}\right]\cdot\nabla_{{\bf r}}g({\bf r})\nonumber \\
 & +\mu_{b}k_{B}T\nabla_{{\bf r}}^{2}g({\bf r})\Bigg\}{\rm d}{\bf r}\nonumber \\
= & \int g({\bf r})\Bigg\{\nabla_{{\bf r}}\cdot\left[\mu_{b}\nabla_{{\bf r}}\left[\int\rho({\bf r}',t)V(|{\bf r}-{\bf r}'|){\rm d}{\bf r}'+U(|{\bf r}-{\bf x_{p}}|)\right]-{\bf f}_{i}-\sqrt{2\mu_{b}k_{B}T}\boldsymbol{\xi}_{i}\right]\rho_{i}({\bf r},t)\nonumber \\
 & +\mu_{b}k_{B}T\nabla_{{\bf r}}^{2}\rho_{i}({\bf r},t)\Bigg\}
\end{align}
In the third step it seems there is an extra term $V(0)$, but it
vanishes due to $\nabla_{{\bf r}}$ operator, and the last step used
part integral. On the other hand, with the identity $\frac{{\rm d}}{{\rm d}t}g({\bf r}_{i})=\frac{{\rm d}}{{\rm d}t}\int\rho_{i}({\bf r},t)g({\bf r}){\rm d}{\bf r}=\int\frac{\partial\rho_{i}({\bf r},t)}{\partial t}g({\bf r}){\rm d}{\bf r}$,
and considering the arbitrariness of function $g({\bf r})$, immediately
\begin{align}
\frac{\partial\rho_{i}({\bf r},t)}{\partial t}= & \nabla_{{\bf r}}\cdot\rho_{i}({\bf r},t)\left[\mu_{b}\nabla_{{\bf r}}\left[\int\rho({\bf r}',t)V(|{\bf r}-{\bf r}'|){\rm d}{\bf r}'+U(|{\bf r}-{\bf x_{p}}|)\right]-{\bf f}_{i}-\sqrt{2\mu_{b}k_{B}T}\boldsymbol{\xi}_{i}\right]\nonumber \\
 & +\mu_{b}k_{B}T\nabla_{{\bf r}}^{2}\rho_{i}({\bf r},t)
\end{align}
then the collective density function 
\begin{align}
\frac{\partial\rho({\bf r},t)}{\partial t}= & \mu_{b}\nabla\cdot\left[\rho({\bf r},t)\nabla\left(\int\rho({\bf r}',t)V(|{\bf r}-{\bf r}'|)+U(|{\bf r}-{\bf x}_{p}|)\right)\right]\nonumber \\
 & +\sum_{i}\left[-\nabla\cdot({\bf f}_{i}\rho_{i})-\sqrt{2\mu_{b}k_{B}T}\nabla\cdot(\rho_{i}\boldsymbol{\xi}_{i})\right]+\mu_{b}k_{B}T\nabla^{2}\rho({\bf r},t)
\end{align}
This equation is not self-consistent yet, since ${\bf f}_{i}\rho_{i}$
and $\rho_{i}\boldsymbol{\xi}_{i}$ terms still exist. To fix this,
following Dean's method \cite{1996_JPA_Dean_LE}, we introduce two
noise fields $\chi_{1,2}({\bf r},t)$ as functions of $\rho({\bf r},t)$
to replace $\chi_{1}'({\bf r},t)=-\sum_{i}\nabla\cdot({\bf f}_{i}\rho_{i})$
and $\chi_{2}'({\bf r},t)=-\sqrt{2\mu_{b}k_{B}T}\sum_{i}\nabla\cdot(\rho_{i}\boldsymbol{\xi}_{i})$.
Considering \begin{subequations}
\begin{align}
\left\langle \chi_{1}'({\bf r},t)\chi_{1}'({\bf r}',t')\right\rangle = & \frac{D_{b}}{\tau_{b}}e^{-|t-t'|/\tau_{b}}\sum_{i}\nabla\cdot\nabla'\left[\rho_{i}({\bf r},t)\rho_{i}({\bf r}',t')\right]\nonumber \\
= & \frac{D_{b}}{\tau_{b}}e^{-|t-t'|/\tau_{b}}\nabla\cdot\nabla'\left[\rho({\bf r},t)\delta({\bf r}-{\bf r}')\right],\\
\left\langle \chi_{2}'({\bf r},t)\chi_{2}'({\bf r}',t')\right\rangle = & 2\mu_{b}k_{B}T\delta(t-t')\sum_{i}\nabla\cdot\nabla'\left[\rho_{i}({\bf r},t)\rho_{i}({\bf r}',t')\right]\nonumber \\
= & 2\mu_{b}k_{B}T\delta(t-t')\nabla\cdot\nabla'\left[\rho({\bf r},t)\delta({\bf r}-{\bf r}')\right],
\end{align}
\end{subequations}we construct noise field $\chi_{1}({\bf r},t)=\nabla\cdot\left[\sqrt{\rho({\bf r},t)}\boldsymbol{\xi}^{A}({\bf r},t)\right]$
and $\chi_{2}({\bf r},t)=\nabla\cdot\left[\sqrt{\rho({\bf r},t)}\boldsymbol{\xi}^{T}({\bf r},t)\right]$
to keep the correlations of $\chi_{1}$ and $\chi_{1}'$, $\chi_{2}$
and $\chi_{2}'$ equal, where $\boldsymbol{\xi}^{A,T}({\bf r},t)$
are also noise field with correlation $\left\langle \boldsymbol{\xi}^{A}({\bf r},t)\boldsymbol{\xi}^{A}({\bf r}',t')\right\rangle =\frac{D_{b}}{\tau}e^{-|t-t'|/\tau_{b}}\delta({\bf r}-{\bf r}'){\bf I}$
and $\left\langle \boldsymbol{\xi}^{T}({\bf r},t)\boldsymbol{\xi}^{T}({\bf r}',t')\right\rangle =2\mu_{b}k_{B}T\delta(-|t-t'|)\delta({\bf r}-{\bf r}'){\bf I}$
respectively. Now we achieve a self-consistent equation for the evolution
of $\rho({\bf r},t)$ 
\begin{align}
\frac{\partial\rho({\bf r},t)}{\partial t}= & \mu_{b}\nabla_{{\bf r}}\cdot\rho({\bf r},t)\nabla_{{\bf r}}\left[\int\rho({\bf r}',t)V(|{\bf r}-{\bf r}'|){\rm d}{\bf r}'+U(|{\bf r}-{\bf x_{p}}|)\right]\nonumber \\
 & +\mu_{b}k_{B}T\nabla^{2}\rho({\bf r},t)+\nabla\cdot\left[\sqrt{\rho({\bf r},t)}\boldsymbol{\xi}^{A}({\bf r},t)\right]+\nabla\cdot\left[\sqrt{\rho({\bf r},t)}\boldsymbol{\xi}^{T}({\bf r},t)\right]\label{eq:Dean}
\end{align}
This equation is one of the central results in this section, also
known as Dean's equation. 

To eliminate variables of bath particle positions, we use a mean-field
theory to describe the active bath. Using Eq.\ref{eq:Dean} and assuming
the environment is isotropic, homogeneous and no special structures(suitable
for weak interaction and dense situations), the evolution equation
for bath density can be simplified as 
\begin{equation}
\frac{\partial\rho_{k}(t)}{\partial t}\approx-\mu_{b}k^{2}\left[(k_{B}T+\rho V_{k})\rho_{k}(t)+\rho U_{k}e^{i{\bf k}\cdot{\bf x}_{p}}\right]+i\sqrt{\rho}{\bf k}\cdot\left[\tilde{\boldsymbol{\xi}}^{T}({\bf k},t)+\tilde{\boldsymbol{\xi}}^{A}({\bf k},t)\right]
\end{equation}
in Fourier space, where $\rho_{k}(t)$, $U_{k}$, $V_{k}$, $\tilde{\boldsymbol{\xi}}^{A}({\bf k},t)$
and $\tilde{\boldsymbol{\xi}}^{T}({\bf k},t)$ are Fourier transforms
of $\rho({\bf r},t)$, $U(r),V(r),\boldsymbol{\xi}^{A}({\bf r},t)$
and $\boldsymbol{\xi}^{T}({\bf r},t)$ respectively. This equation
has a formal solution 
\begin{equation}
\rho_{k}(t)=\int_{-\infty}^{t}e^{-(t-s)/a_{k}}\left[-\mu_{b}k_{B}Tk^{2}\rho U_{k}e^{i{\bf k}\cdot{\bf x}_{p}}+i\sqrt{\rho}{\bf k}\cdot(\tilde{\boldsymbol{\xi}}^{A}({\bf k},s)+\tilde{\boldsymbol{\xi}}^{T}({\bf k},s))\right]{\rm d}s\label{eq:rho_k_form}
\end{equation}
where $a_{k}=\left[\mu_{b}k^{2}(k_{B}T+\rho V_{k})\right]^{-1}$.
Using the identity (performing Fourier transition and its inverse transform on the l.h.s.)
\begin{equation}
-\nabla_{{\bf x}_{p}}\sum_{i}U(|{\bf r}_{i}-{\bf x}_{p}|)\equiv\frac{1}{(2\pi)^{3}}\int i{\bf k}e^{-i{\bf k}\cdot{\bf x}_{p}}\rho_{k}(t) U_k {\rm d}^{3}{\bf k},
\end{equation}
and inserting the formal solution (\ref{eq:rho_k_form}) into the Langevin equation for probe Eq.(1) in main text, we get a generalized Langevin equation for probe movement along $\vec{e}_{x}$-direction.
\begin{equation}
\dot{x}_{p}=\mu_{p}\int_{-\infty}^{t}\tilde{F}(t-s){\rm d}s+\mu_{p}[K(x_{m}-x_{p})+F_{0}]+\eta_{A}(x_{p}(t),t)+\eta_{T}(x_{p}(t),t)+\sqrt{2\mu_{p}k_{B}T}\xi_{t}
\end{equation}
where $\tilde{F}(t)=-\frac{\mu_{b}\rho}{(2\pi)^{3}}\int ik_{x}k^{2}U_{k}^{2}e^{-ik_{x}[x_{p}(t)-x_{p}(s)]}e^{-(t-s)/a_{k}}$
is a complex memory kernel, and 
\begin{equation}
\eta_{A,T}(x_{p}(t),t)=\frac{\mu_{p}\sqrt{\rho}}{(2\pi)^{3}}\int ik_{x}U_{k}e^{-ik_{x}x_{p}(t)}\int_{-\infty}^{t}e^{-(t-s)/a_{k}}{\bf k}\cdot\tilde{\boldsymbol{\xi}}^{A,T}({\bf k},s){\rm d}s{\rm d}^{3}{\bf k}
\end{equation}
is the colored noise term induced by bath. This memory kernel is far
complex to use, yet to the linear order, the memory kernel can be
simplified to the form $\mu_{p}\int_{-\infty}^{t}\tilde{F}(t-s){\rm d}s\approx-\int_{-\infty}^{t}\zeta(t-s)\dot{x}_{p}(s){\rm d}s$,
where 
\begin{equation}
\zeta(t)=\frac{\mu_{p}\mu_{b}\rho}{3(2\pi)^{3}}\int k^{4}U_{k}^{2}a_{k}e^{-t/a_{k}}{\rm d}^{3}{\bf k}
\end{equation}
which is much easier to employ. As for the noise $\eta_{A,T}(x_{p}(t),t)$,
considering the time scale of probe movement is much slower than bath
particles, we use the adiabatic approximation so that the noises can
be simplified into 
\begin{equation}
\eta_{A,T}(t)=\frac{\mu_{p}\sqrt{\rho}}{(2\pi)^{3}}\int ik_{x}U_{k}\int_{-\infty}^{t}e^{-(t-s)/a_{k}}{\bf k}\cdot\tilde{\boldsymbol{\xi}}^{A,T}({\bf k},s){\rm d}s{\rm d}^{3}{\bf k}\label{eq:eta_AT}
\end{equation}
with time correlations \begin{subequations}
\begin{align}
\left\langle \eta_{T}(t)\eta_{T}(t')\right\rangle  & =\frac{2\mu_{t}^{2}\rho\mu_{b}k_{B}T}{(2\pi)^{3}}\int k_{x}^{2}U_{k}^{2}k^{2}\frac{a_{k}}{2}e^{-|t-t'|/a_{k}}{\rm d}^{3}{\bf k}\\
\left\langle \eta_{A}(t)\eta_{A}(t')\right\rangle  & =\frac{\mu_{t}^{2}\rho D_{b}}{(2\pi)^{3}}\int k_{x}^{2}U_{k}^{2}k^{2}\frac{1}{(\tau_{b}/a_{k})^{2}-1}\left[\tau_{b}e^{-|t-t'|/\tau_{b}}-a_{k}e^{-|t-t'|/a_{k}}\right]{\rm d}^{3}{\bf k}
\end{align}
\label{eq:corr_eta}
\end{subequations}

\section{Generation of Complex Colored Noise} \label{sec:eta}

According to Eq.\eqref{eq:eta_AT}, and using Greek alphabet to express vector component,  $\boldsymbol{\eta}_{A,T}$ in $\alpha$ component is  
\begin{equation}
    \eta_{A,T}^\alpha (t) = \frac{\mu_t \sqrt{\rho}}{(2\pi)^{3} } \int \dif^3 \bk k^\alpha e^{-i \bk \cdot {\bf x}_p(t)} U_k \int_{-\infty}^{t} e^{-(t-s)/a_k} [ \sum_{\beta} k^\beta \tilde{\xi}^{A,T}_\beta (\bk, s)] {\rm d}s \label{eq:etaAT_SI}
\end{equation}Since $\tilde{\xi}_{\beta}^{A,T}(\bk,t) = \int {\xi}^{A,T}_{\beta} ({\bf r},t) e^{i\bk \cdot \br} \dif^3 \br$, as well as the correlations shown in Sec.\ref{sec:dean}, one has
\begin{subequations}
\begin{align}
    \bka{\tilde{\xi}^{A*}_\alpha (\bk,t) \tilde{\xi}^{A}_\beta (\bk', t')} &= \frac{D_b}{\tau_b} \delta_{\alpha \beta} (2\pi)^3 \delta(\bk - \bk') e^{-|t-t'|/\tau_b} \\
    \bka{\tilde{\xi}^{T*}_\alpha (\bk,t) \tilde{\xi}^{T}_\beta (\bk', t')} &= 2\mu_b k_B T \delta_{\alpha \beta} (2\pi)^3 \delta(\bk - \bk') \delta(t-t')
\end{align}
\end{subequations}
Therefore random variables $\tilde{\xi}^{A,T}_\alpha({\bf k},t)$ can be devided into two independent stochastic processes in time and $k$ space,
\begin{subequations}
\begin{align}
    \tilde{\xi}^A_\alpha (\bk, t) \dif t \dif^3 \bk &= \sqrt{\frac{D_b}{\tau_b}}(2\pi)^{3/2} f_\alpha(t) \dif t \dif^3 W_{\bk} \\
    \tilde{\xi}^T_\alpha (\bk, t) \dif t \dif^3 \bk &= \sqrt{2\mu_b k_B T} (2\pi)^{3/2} \dif W_t \dif^3 W_{\bk}
\end{align}
\end{subequations}
where $W_t$ and $W_{\bk}$ are independent Wiener processes, $f_\alpha(t)$ is an dimensionless OU process with $\tau_b \dot{f}_\alpha(t) = -f_\alpha (t) + \sqrt{2\tau_b} \xi_t $ ($\xi_t$ stands for standard white noise), formal solution $f_\alpha(t) = \sqrt{\frac{2}{\tau_b}} \int_{-\infty}^t e^{-(t-s)/\tau_b} \xi_s \dif s$ and time correlation $\bka{f_\alpha(t) f_\beta(t')} = \delta_{\alpha\beta} e^{-|t-t'|/\tau_b}$. 

This proposal indicates Eq.\eqref{eq:etaAT_SI} can be rewritten as
\begin{subequations}
\begin{align}
    \eta_A^\alpha(t) &= -\frac{\mu_t}{(2\pi)^{3/2}} \sqrt{\rho D_b} \int k_\alpha e^{-i\bk \cdot {\bf x}(t) }U_k  \sum_{\beta} k_\beta B^A_\beta(\bk, t)  \dif^3 W_{\bk} \nonumber \\
    &= -\frac{\mu_t}{d(2\pi)^{3/2}} \sqrt{\rho D_b} \int k^2 e^{-i\bk \cdot {\bf x}(t) }U_k B^A_\alpha(\bk, t)  \dif^3 W_{\bk} \\
    \eta_T^\alpha (t) &= -\frac{\mu_t}{(2\pi)^{3/2}} \sqrt{2\rho \mu_b k_B T} \int k_\alpha e^{-i\bk \cdot {\bf x}(t) }U_k  \sum_{\beta} k_\beta B^T_\beta(\bk, t)   \dif^3 W_{\bk} \nonumber \\
    &= -\frac{\mu_t}{d(2\pi)^{3/2}} \sqrt{2\rho \mu_b k_B T} \int k^2 e^{-i\bk \cdot {\bf x}(t) }U_k B^T_\alpha (\bk, t)  \dif^3 W_{\bk} 
\end{align}
\end{subequations}
where $B^A_\alpha(\bk,t) = \int_{-\infty}^t \tau_b^{-1/2} e^{-(t-s)/a_k} f_\alpha(s) \dif s$, $B^T_\alpha (\bk,t) =  \int_{-\infty}^t e^{-(t-s)/a_k} \dif W_s $ are independent stochastic processes which can be generated numerically. 

In detail, one has $\dot{B}_\alpha^T = -a_k^{-1} B_\alpha^T + \xi_t$,  which is also an OU process with $\bka{B^T_\alpha (t) B^T_\alpha(t')} = \frac{a_k}{2} e^{-|t-t'|/a_k}$. Therefore the initial value of $B_\alpha^T(\bk, t)$ can be set as a Gaussian random number with zero mean and variance $a_k/2$. Numerically, $B_\alpha^T(\bk, t_{i+1}) = e^{-\Delta t / a_k} B_\alpha^T(\bk, t_i) + \sqrt{\frac{a_k}{2} (1-e^{-2\Delta t/a_k}) }u_{i+1}$, where $\Delta t=t_{i+1} - t_{i}$ is the time interval, $\{u_{i}\}$ is a set of independent Gaussian random variables of zero mean and variance 1. Here we emphasize that the ordinary Eular-Maruyama algorithm (i.e. $B_\alpha^T(\bk,t_{i+1}) = (1-\Delta t /a_k) B_\alpha^T(\bk, t_i) + \sqrt{\Delta t} u_i$) is not suitable for the present case, since the characteristic time scale $a_k$ is dependent on $k=|\bk|$ and it is not practical to choose a small enough interval s.t. $\Delta t \ll a_k$ for all $k$s. 

On the other hand, $\dot{B}_\alpha^A = -a_k^{-1} B_\alpha^A + \tau_b^{-1/2} f_\alpha(t)$ and $\ddot{B}_\alpha^A + (a_k^{-1}+\tau_b^{-1}) \dot{B}_\alpha^A + (a_k \tau_b)^{-1} B_\alpha^A = \frac{\sqrt{2}}{\tau_b} \xi_t$, leads to the solution
\begin{equation}
B_\alpha^A(\bk, t)  = \frac{a_k \sqrt{2}}{\tau_b - a_k } \int_{-\infty}^t \bks{e^{-(t-s)/\tau_b} - e^{-(t-s)/a_k}} \dif W_s
\end{equation}
with correlation function $\bka{B^A_\alpha (t) B_\alpha^A(t')} = \frac{1}{(\tau_b/a_k)^2 - 1}  \bks{\tau_b e^{-|t-t'|/\tau_b} - a_k e^{-|t-t'|/a_k} }$. So the inital value of $B_\alpha^A(t)$ can be set of Gaussian random number with zero mean and variance $\frac{a_k^2}{\tau_b + a_k}$. Consequently, $B_\alpha^A(\bk,t)$ can be written as 
\begin{equation}
B_\alpha^A(\bk,t_{i+1}) =e^{-\Delta t/a_k} B_\alpha^A(\bk,t_{i}) + \frac{a_k\sqrt{2}}{\tau_b - a_k} \bks{ (e^{-\Delta t/\tau_b} - e^{-\Delta t/a_k}) \sqrt{\frac{\tau_b}{2}}f_\alpha(t_i) + G_\alpha(\bk,t_i)},
\end{equation}
where $ G_\alpha(\bk,t_i) = \int_{t_i}^{t_{i+1}} \bks{e^{-(t_{i+1} -s)/\tau_b} - e^{-(t_{i+1} - s)/a_k} } \dif W_s$, with expectation 
\begin{align}
    \bka{G_\alpha(\bk,t_i) G_\alpha(\bk,t_j) } &= \delta_{ij} \bks{\frac{\tau_b}{2}(1-e^{-2\Delta t/\tau_b}) + \frac{a_k}{2} (1-e^{-2\Delta t/a_k}) - \frac{2\tau_b a_k}{a_k+\tau_b} (1 - e^{-\Delta t/\tau_b} e^{-\Delta t/a_k})} \nonumber \\
    & \equiv \delta_{ij} \mathcal{G}_\alpha(\bk,\Delta t)
\end{align}
which is in order $\frac{\Delta t^3 (a_k-\tau_b)^2}{3a_k^2\tau_b^2} + O(\Delta t^4)$. Finally, the exact numerical algorithm to generate $B_\alpha^A(\bk, t_i)$ is
\begin{subequations}
\begin{align}
    B_\alpha^A(\bk,t_{i+1}) &= e^{-\Delta t/a_k} B_\alpha^A(\bk,t_{i}) + \frac{a_k\sqrt{2}}{\tau_b - a_k} \bks{ (e^{-\Delta t/\tau_b} - e^{\Delta t/a_k}) \sqrt{\frac{\tau_b}{2}}f_\alpha(t_i) + \sqrt{\mathcal{G}_\alpha(\bk,\Delta t)} v_i}\\
    f_\alpha (t_{i+1}) &= e^{-\Delta t/\tau_b} f_\alpha (t_i) + \sqrt{1-e^{-2\Delta t/\tau_b} } w_i
\end{align}
\end{subequations}
where $\{v_i\}$ and $\{w_i\}$ are sets of independent Gaussian random variables of zero mean and variance 1. Comparing with the direct differential algorithm $B_\alpha^A (t_{i+1})= (1-\Delta t/a_k) B_\alpha^A(t_i) + \tau_b^{-1/2} f_\alpha (t) \Delta t $, our method is suitable for the situation when $\Delta t \ge a_k$.

For 3d system, stochastic integral over $\dif^3 W_{\bk}$ can be simplified through following method. For simplicity, consider an arbitrary bounded stochastic integral $\eta = \iiint f(\bk) \dif^3 W_\bk$ with $f(\bk) = f(k)$, $k=|\bk|$. One has $\bka{\eta} = 0$ and $\bka{\eta^2} = \iiint f^2(k) \dif^3 \bk = 4\pi \int_0^\infty f^2(k) k^2 \dif k$. Now consider another one-dimensional integral $\varphi = a\int_0^\infty f(k) k^b \dif W_k$, one also has $\bka{\varphi^2} = a^2\int_0^\infty f^2(k) k^{2b} \dif k$. Let $\bka{\eta^2} = \bka{\varphi^2}$, immediately one gets $a=\sqrt{4\pi}$ and $b = 1$. This method can greatly simplify the calculation of $\eta_{A,T}^\alpha({\bf x},t)$.

At last, under the adiabatic approximation, we have
\begin{subequations}
\begin{align}
    \eta_T^\alpha (0,t) &= -\frac{\mu_t}{3(2\pi)^{3/2}} \sqrt{8 \pi \rho \mu_b k_B T} \int_0^{\infty} k^3 U_k B_\alpha^T(k,t) \dif W_k \\
    \eta_A^\alpha (0,t) &= -\frac{\mu_t}{3(2\pi)^{3/2}} \sqrt{4 \pi \rho D_b} \int_0^{\infty} k^3 U_k B_\alpha^A(k,t) \dif W_k
\end{align}
\end{subequations}

Now we consider the asymptotic behavior of correlation function at large time scale. In this situation, only very small $k$s contribute to the intergral. Therefore one may assume $e^{-|t-t'|/a_k} \approx e^{-|t-t'|\mu_b k^2 (k_BT + \rho V_{0})}$, where $V_0$ notes for $V(\bk=0)$. Consequently, for $\eta_T$, one has
\begin{align}
     \bka{\eta_T^\alpha(0,t) \eta_T^\alpha(0,t')} \approx & \frac{\mu_t^2 \rho \mu_b k_B T}{(2\pi)^d} \int \frac{k_\alpha^2 U_0^2}{\mu_b (k_B T + \rho V_0)} e^{-|t-t'|\mu_b (k_B T + \rho V_0) k^2} \dif^3 \bk \nonumber \\
     =&\frac{\mu_t^2 \rho }{(2\pi)^d (1+\rho V_0/k_B T)} \frac{1}{2p}  \bkr{\frac{\pi}{p}}^{\frac{d}{2}} 
\end{align}
where $U_0$ also stands for $U(\bk=0)$,  $p = |t-t'|\mu_b (k_B T + \rho V_0)$. As a result, we get $\bka{\eta_T^\alpha(0,t) \eta_T^\alpha(0,t')} \propto |t-t'|^{-(d/2+1)}$. 

For $\eta_A$, the exponential decay part $e^{-|t-t'|/\tau_b}$ has no contribution to the long time decay behavior anyway. One may only consider the other part, i.e.
\begin{align}
    \bka{\eta_A^\alpha(0,t) \eta_A^\alpha(0,t')} \asymp & \frac{\mu_t^2 \rho D_b}{(2\pi)^2} \int k_\alpha^2 U_k^2 \frac{k^2 a_k}{1-(\tau_b / a_k)^2} e^{-|t-t'|/a_k} \dif^3 \bk \nonumber \\
    \approx & \frac{\mu_t^2 \rho }{(2\pi)^d} \frac{D_b}{\mu_b (k_B T + \rho V_0)} \int \frac{k_\alpha^2 U_0^2 }{1 - (\mu_b (k_B T + \rho V_0) \tau_b)^2 k^4} e^{-|t-t'|\mu_b (k_B T + \rho V_0) k^2 }  \dif^3 \bk \nonumber \\
    \approx &  \frac{\mu_t^2 \rho }{(2\pi)^d} \frac{D_b}{\mu_b (k_B T + \rho V_0)}  \frac{1}{2p}  \bkr{\frac{\pi}{p}}^{\frac{d}{2}} 
\end{align}
Clearly the long-time behavior also follows a power law $ \bka{\eta_A^\alpha(0,t) \eta_A^\alpha(0,t')} \propto |t-t'|^{-(d/2+1)}$.

Another situation is the weak interaction limit between bath particles, $k_BT \gg \rho V(k)$, then $a_k \approx \bks{\mu_b k^2 k_BT}^{-1}$. Herein the correlation of $\eta_T$ is 
\begin{align}
    \bka{\eta_T^\alpha(0,t) \eta_T^\alpha(0,t')} & \approx \frac{\mu_t^2 \rho \mu_b k_B T}{(2\pi)^d} \int \frac{k_\alpha^2 \iint e^{i\bk \cdot (\br + \br')} U(r)U(r') \dif \br \dif \br'}{\mu_b k_B T} e^{-|t-t'|\mu_b k_B T k^2} \dif^3 \bk \nonumber \\
    &=\frac{\mu_t^2 \rho}{(2\pi)^2} \int k_\alpha^2 \iint  U(r)U(r') e^{-|t-t'|\mu_b k_B T k^2 + i\bk \cdot (\br + \br')} \dif \br \dif \br' \dif^3 \bk \nonumber \\
    &= \frac{\mu_t^2 \rho }{(2\pi)^d } \iint \bks{\frac{1-2q^2_\alpha}{p} e^{-{\bf q}^2}} \bkr{\frac{\pi}{p}}^{\frac{d}{2}} U(r)U(r') \dif \br \dif \br'
\end{align}
where $p = |t-t'|\mu_b k_B T$, ${\bf q} = \frac{(\br+\br')}{2\sqrt{p}}$. When $|t-t'|$ is large enough, $\bka{\eta_T^\alpha(0,t) \eta_T^\alpha(0,t')} \propto |t-t'|^{-(d/2+1)}$

\end{widetext}

\bibliographystyle{apsrev4-2}

\end{document}